\documentclass[conference,10pt]{IEEEtran}
\usepackage{epsfig,rotating,setspace,latexsym,amsmath,epsf,amssymb,amsfonts,bm,theorem,subfigure,epstopdf}
\usepackage{cite,authblk}
\usepackage{bbm}
\usepackage{algorithmic,algorithm}

\newtheorem{lemma}{Lemma}
\newenvironment{Proof}[1]{\medskip\par\noindent{\bf Proof:\,}\,#1}{{\mbox{\,$\blacksquare$}\par}}
\IEEEoverridecommandlockouts

\allowdisplaybreaks
\usepackage{color}

\begin{document}

\title{Scheduling a Human Channel\thanks{This work was supported by NSF Grants CNS 15-26608, CCF 17-13977 and ECCS 18-07348.}}

\author{Melih Bastopcu \qquad Sennur Ulukus\\
	\normalsize Department of Electrical and Computer Engineering\\
	\normalsize University of Maryland, College Park, MD 20742\\
	\normalsize  \emph{bastopcu@umd.edu} \qquad \emph{ulukus@umd.edu}}

\maketitle

\begin{abstract}	
We consider a system where a human operator processes a sequence of tasks that are similar in nature under a total time constraint. In these systems, the performance of the operator depends on its past utilization. This is akin to \emph{state-dependent} channels where the past actions of the transmitter affects the future quality of the channel (also known as \emph{action-dependent} or \emph{use-dependent} channels). For \emph{human channels}, a well-known psychological phenomena, known as \emph{Yerkes-Dodson law}, states that a human operator performs worse when he/she is over-utilized or under-utilized. Over such a \emph{use-dependent} human channel, we consider the problem of maximizing a utility function, which is monotonically increasing and concave in the time allocated for each task, under explicit minimum and maximum \emph{utilization} constraints. We show that the optimal solution is to keep the utilization ratio of the operator as high as possible, and to process all the tasks. We prove that the optimal policy consists of two major strategies: utilize the operator without resting until reaching the maximum allowable utilization ratio, and then alternate between working and resting the operator each time reaching the maximum allowable utilization at the end of work-period. We show that even though the tasks are similar in difficulty, the time allocated for the tasks can be different depending on the strategy in which a task is processed; however, the tasks processed in the same strategy are processed equally.
\end{abstract}

\section{Introduction}
We consider scheduling a human operator who processes a sequence of tasks that are similar in difficulty over a fixed duration. Performance of a human operator is closely related to his/her workload. Yerkes-Dodson law \cite{yerkes_dodson} states that human operators perform worse when their workload is too high or too low. We use \emph{utilization ratio} to keep track of the operator's past workload. Utilization ratio of the operator increases when the operator processes a task, and decreases when the operator idles (rests). We enforce explicit constraints in the form of minimum and maximum allowable utilization ratios, in order to keep the performance of the human operator high.

This problem is intimately related to a general class of problems in communication and information theory. In the communication and information theoretic treatment of certain modern applications, the channel can no longer be modeled as static or i.i.d.~over time. In such applications, the characteristics of the communication channel changes as a function of its past utilization. Examples include, for instance, channel that die \cite{channels_that_die}, channels that heat up \cite{channels_that_heat_up, omur_that_heat_up, omur_abdulrahman_that_heat_up}, channels that wear out over time \cite{channel_that_wears_out}, binary symmetric channel where the crossover probability changes over time due to usage \cite{Use-Dependent_Packet-Drop_Channels}, channels that get biased over time \cite{david_ward}, and queuing systems where the service quality of the queue depends on the queue length \cite{Queue_Length_Dependent_Service}. In particular, we will see a remarkable similarity between scheduling a human operator subject to utilization ratio constraints in this paper and scheduling a communication channel subject to temperature constraints in \cite{omur_that_heat_up, omur_abdulrahman_that_heat_up}.

When the operator processes a task, the system receives a certain reward (utility). We model the utility function, $u(t)$, as a monotonically increasing concave function of the processing time. Examples of such utility functions are observed, for instance, in speed accuracy trade-off (SAT) studies \cite{SAT_curves} where the utility function is modeled as an exponential growth to a saturation point, as $u(t)=1-e^{-at}$; in rate-distortion \cite[Eqns. (1), (3), (4)]{arafa-isit2017} where the time required to achieve an outcome with a certain distortion under a fixed energy is $u(t)=\frac{a}{1+b/t}$; and in many scenarios where more time spent on a task results in diminishing (sub-linear) returns over time, e.g., a runner can make considerable improvement at the initial stages of training, a student can quickly answer easier parts of the questions, a monitor can quickly determine the general area where a target is, however, in each of these examples, getting a higher running performance, solving difficult parts of the questions, detecting the target with more precision, require much more time, and returns become sub-linear. Another simple such sub-linear function is $u(t) = \log(1+t)$.

Our work is most closely related to \cite{Humans_and_UAVs, dynamical_queue, task_release, adaptive_attention}. In \cite{adaptive_attention}, the authors consider sigmoid functions for the utility function, whereas here, we consider monotone increasing concave functions. \cite{adaptive_attention} imposes minimum processing times for the tasks, prioritizes the tasks, and considers the case where some tasks are mandatory. In our paper, there is no minimum time allocation for the tasks, and all the tasks are identical in importance and difficulty. Thus, our model can be viewed as a simplified version of \cite{adaptive_attention}. Our goal for this simplification is to obtain general and structural results, as we discuss next.

In this paper, we consider a scheduling problem for a single human operator who performs tasks similar in difficulty over a given fixed time. The number of tasks $N$ and the total duration $T$ are known a priori. The structural solution for this problem consists of two major sub-policies: In the first policy, the operator starts processing tasks, and continues to process tasks without idling until he/she reaches the allowable upper bound for his/her utilization ratio. In the second policy, which starts after the operator reaches the allowable upper bound for the utilization ratio, the operator must idle (rest) during each task. We show that the operator should allocate equal time for each task it performs in the same sub-policy. However, the times allocated for the tasks performed during different sub-policies may be different even though the tasks are identical in difficulty. We note that the structure of the utilization ratio here is similar to the evolution of the temperature in the case of single energy arrival in \cite{omur_that_heat_up}.

\section{System Model and the Problem}
\begin{figure}[t] \centerline{\includegraphics[width=1\columnwidth]{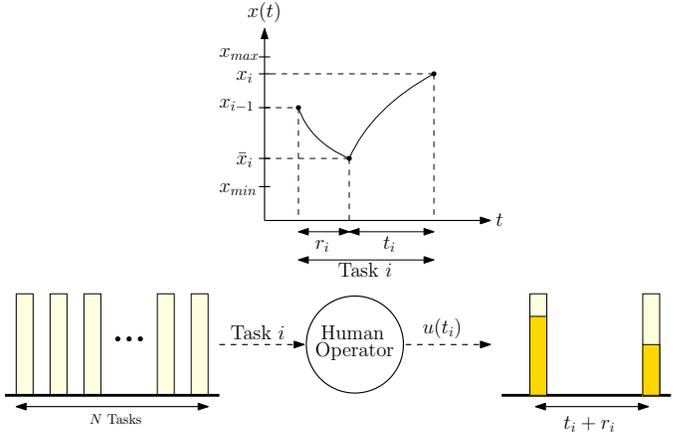}}
	\caption{A human operator processing $N$ tasks in a duration of $T$. For task $i$, the operator idles for $r_i$, processes for $t_i$, and receives an award of $u(t_i)$.}
	\label{sys_model}
	\vspace{-0.4cm}
\end{figure}

We consider a system where a human operator processes $N$ tasks over a duration of $T$ units of time, see Fig.~\ref{sys_model}. We model the \emph{utilization ratio}, $x(t)$, where $x(t) \in [0, 1]$, as \cite{adaptive_attention},
\begin{align}
\frac{d x(t)}{dt} & = \alpha (b(t)-x(t)), \qquad  x(0)=x_{0} \label{eq1}
\end{align}
where $b(t)=1$ if the operator is working at time $t$, and $b(t)=0$ if the operator is idling at time $t$, and $\alpha$ (which is denoted as
$\frac{1}{\delta}$ in \cite{adaptive_attention}) is a constant that depends on the resistance of the operator to the workload.\footnote{Note the similarity between utilization-workload equation in (\ref{eq1}) and temperature-power equation in \cite[Eqn. (3)]{omur_that_heat_up}. In particular, (\ref{eq1}) here is the same as \cite[Eqn. (3)]{omur_that_heat_up} when $T_e=0$, $c=0$ and $a=b=\alpha$ in \cite[Eqn. (3)]{omur_that_heat_up}, with the mapping of utilization ratio ($x(t)$) and workload ($b(t)$) here to temperature ($T(t)$) and transmit power ($P(t)$), respectively, in \cite{omur_that_heat_up}.} Here, $x(t)$ increases when the operator is working, and decreases when the operator is idling. After resting for $r$ and working for $t$, $x(t)$ evolves as,
\begin{align}
x(t+r) = 1-e^{-\alpha t} +x_{0}e^{- \alpha (r+t)}
\end{align}

According to the Yerkes-Dodson law \cite{yerkes_dodson}, the performance of the operator will be worse if the utilization ratio $x(t)$ is too low or too high. Therefore, we aim to keep $x(t)$ between a pre-specified minimum, $x_{min}$, and maximum, $x_{max}$. For each task $i$, the operator works (processes the task) for $t_i$ seconds and rests (idles) for $r_i$ seconds. Without loss of generality, we assume that the operator idles first (if any) before processing a task. We denote by $x_i$ the utilization ratio $x(t)$ right after the operator finishes processing task $i$. We denote by $\bar{x}_i$ the utilization ratio $x(t)$ right before the operator starts processing task $i$, i.e., right after the operator finishes resting (if any) for task $i$; see the top part of Fig.~\ref{sys_model}. Thus,
\begin{align}
\bar{x}_i &= x_{i-1}e^{-\alpha r_i} \\
x_i &= 1-e^{-\alpha t_i}+\bar{x}_i e^{-\alpha t_i}
\end{align}

Due to the monotonicity of $x(t)$ during processing and idling periods, if the initial utilization ratio is between $x_{min}$ and $x_{max}$, it suffices to check the utilization ratio only at the ends of idling and processing times, i.e., at $\bar{x}_i$ and $x_i$, to make sure that it is between $x_{min}$ and $x_{max}$ at all times. The reward acquired from task $i$ is $u(t_i)$. Thus, we formulate the problem,
\begin{align}
\label{problem_1}
\max_{\{t_{i}, r_{i} \}}  \quad &  \sum_{i=1}^{N} u \left(t_{i}\right) \nonumber \\
\mbox{s.t.} \quad & \sum_{i=1}^{N} t_{i}+r_{i} \leq T  \nonumber \\
 & \bar{x}_i \geq x_{min}, \quad  x_i\leq x_{max}, \quad \forall i
\end{align}
which we solve in the rest of this paper.

\section{Structure of the Optimal Solution} \label{sect:structure}
In this section, we identify some important properties of the optimal solution for the problem given in (\ref{problem_1}). First, the following lemma states that, in the optimal solution, if the total time $T$ is not completely utilized, then the operator must have hit the minimum and maximum allowable utilization ratios for every task by resting and working as much as possible.

\begin{lemma}\label{lemma_1}
In the optimal policy, if $\sum_{i=1}^{N} t_{i}+r_{i} < T$, then $\bar{x}_i = x_{min}$ and $x_i = x_{max}$, for all $i$.
\end{lemma}

\begin{Proof}
We prove this by contradiction. Assume that $\sum_{i=1}^{N} t_{i}+r_{i} < T$ and one of the following cases is true: i) $\bar{x}_i > x_{min}$ and $x_i = x_{max}$, ii) $\bar{x}_i= x_{min}$ and $x_i<x_{max}$, or iii) $\bar{x}_i > x_{min}$ and $x_i < x_{max}$. Consider case i). Since the total time constraint is inactive, we can increase $r_i$ without violating any other constraints, and then, increase the corresponding $t_i$. The resulting new policy gives strictly higher reward. In this case, we can increase the reward until either $\sum_{i=1}^{N} t_{i}+r_{i} = T$, or $\sum_{i=1}^{N} t_{i}+r_{i} < T$ and $\bar{x}_i = x_{min}$. Thus, if $\sum_{i=1}^{N} t_{i}+r_{i} < T$, then $\bar{x}_i > x_{min}$ cannot be optimal. In case ii), we can increase $t_i$ and $r_{i+1}$ so that the policy is still feasible and gives higher reward. We can continue to increase $t_i$ until either $\sum_{i=1}^{N} t_{i}+r_{i} = T$, or $\sum_{i=1}^{N} t_{i}+r_{i} < T$ and $x_i = x_{max}$. Thus, if  $\sum_{i=1}^{N} t_{i}+r_{i} < T$, then $x_i = x_{max}$ cannot be optimal. In case iii), we can apply the procedure in ii) first to make $x_i = x_{max}$, which will bring the setting to the case in i), and we can apply the process in i) next. 	
\end{Proof}

Therefore, in the remainder, we focus on the case where the allowed time $T$ is completely utilized. Then, at time $T$, the utilization ratio $x(T)$ will either reach its maximum allowed value $x_{max}$ or not. The following lemma identifies the optimal solution when the utilization ratio does not reach $x_{max}$ at $T$.

\begin{lemma}\label{lemma_2}
In the optimal policy, when $\sum_{i=1}^{N} t_{i}+r_{i} = T$: if $x(T) < x_{max}$, then $r_i = 0$ and $ t_i = \frac{T}{N}$, for all $i$.
\end{lemma}

\begin{Proof}
We prove this by contradiction. Assume that $x(T) < x_{max}$ and $r_i\neq 0$ for some $i$. Choose the maximum task index, $j$, such that $r_j\neq 0$. Since the operator idles before processing a task, the operator completes the remaining tasks after idling for $r_j$, and without idling for the rest of the tasks. Since $x(T) < x_{max}$, we can decrease $r_j$ and increase the processing times of the remaining tasks. The new policy is still feasible and gives a larger reward. We continue to apply this process until either $r_j = 0$ or $x(T) = x_{max}$. If $x(T)<x_{max}$ and $r_j = 0$, then we choose the next highest task index $k$ such that $r_k\neq 0$ and apply the same procedure. At the end, either $x(T) = x_{max}$ or if  $x(T) < x_{max}$, then $r_i = 0$, for all $i$. Thus, if at the end $x(T) < x_{max}$, we have all $r_i=0$ in this case, and the transition from $x_0$ to $x(T)$ can be expressed as,
\begin{align}
x(T) = 1-e^{-\alpha T}+x_0e^{-\alpha T}
\end{align}
Note that in this case, transition of utilization ratio is independent of the time allocated to each task. Since the reward function is a symmetric sum of concave functions, allocating equal amount of time for each task gives the highest reward. Thus, $t_i = \frac{T}{N}$ for all $i$ is optimal, if $x(T) < x_{max}$.
\end{Proof}

Thus, in the remainder, we focus on the case where the allowed time $T$ is completely utilized and the utilization ratio at the end reaches $x_{max}$, that is $x(T)=x_{max}$. The following lemma states that, in this case, if the operator does not reach $x_{max}$ for a task, then he/she should not idle for that task.

\begin{figure}[t]
	\centerline{\includegraphics[width=0.95\columnwidth]{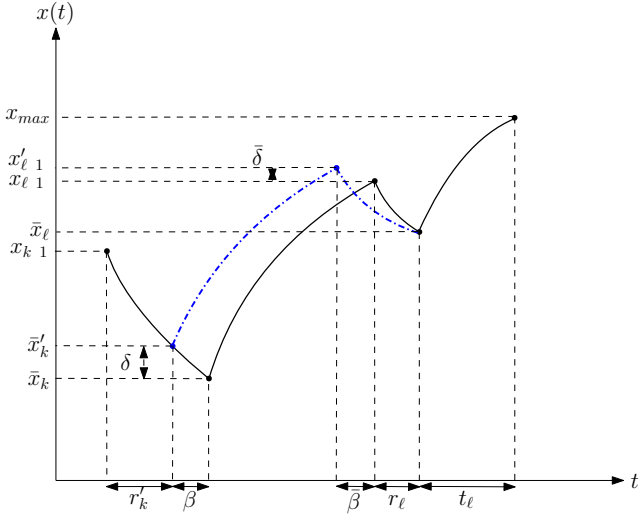}}
	\caption{Black curve shows the hypothetical optimal policy; blue dashed curve shows the new feasible policy that yields a larger reward (Lemma~\ref{lemma_3}).}
	\label{lemma3_fig}
\vspace*{-0.4cm}
\end{figure}

\begin{lemma}\label{lemma_3}
In the optimal policy, when $\sum_{i=1}^{N} t_{i}+r_{i} = T$ and $x(T)=x_{max}$: for any given task $i$, if $x_i< x_{max}$, then $r_i = 0$.
\end{lemma}

\begin{Proof}
We prove this by contradiction. Assume that there is an optimal policy such that there exists an index $i$ where $r_i \neq 0$ and $x_i\neq x_{max}$. From Lemma~\ref{lemma_2}, we know that if $r_i\neq 0$ for some $i$, then $x(T) = x_{max}$. Thus, $x_i = x_{max}$  is satisfied at least once at $i=N$. Let $k$ be the largest $i$ such that $r_i \neq 0$ and $x_i\neq x_{max}$, and choose the smallest $\ell$ such that $\ell>k$ and $x_\ell = x_{max}$. We know that $\ell$ exists since $\ell=N$ satisfies the condition. Then, we construct a new feasible policy such that the difference of $x_{k-1}-\bar{x}_k$ is decreased by $\delta $ and the difference of $x_{\ell-1}-\bar{x}_\ell$ is increased by $\bar{\delta}$, by decreasing the resting time of task $k$. We denote the new policy with primes. The original and new policies are shown in Fig.~\ref{lemma3_fig}. Then,
\begin{align}
\delta &=x_{k-1}e^{-\alpha r_{k}'}-x_{k-1}e^{- \alpha r_{k}}  \\
 \bar{\delta} &= x_{\ell-1}'-x_{\ell-1}'e^{-\alpha \bar{\beta}}
\end{align}
where $r_{k}'$ is the resting time for the $k$th task in the new policy, and $r_{k}-r_k' = \beta$, $r_{\ell}'-r_\ell = \bar{\beta}$. Since $\bar{x}_k$ is increased by $\delta$, $x_{\ell-1}$ is also changed to be $x_{\ell-1}'$ where $x_{\ell-1}'-x_{\ell-1} = \bar{\delta}$. Then,
\begin{align}
 \bar{\delta} &= x_{\ell-1}' -  x_{\ell-1} =\delta e^{-\alpha\sum_{i=l}^{j-1} t_i} \\
 x_{k-1}e^{-\alpha r_{k}'} (1-e^{-\alpha \beta}) &= \delta > \bar{\delta} = x_{\ell-1}'(1-e^{-\alpha \bar{\beta}})
\end{align}
Since $x_{k-1}e^{-\alpha r_{k}'} < x_{\ell-1}'$, we have $1-e^{-\alpha \beta}> 1-e^{-\alpha \bar{\beta}}$ which implies $\bar{\beta} < \beta$. Thus, we can decrease the time for idling by an amount of $ \beta- \bar{\beta}> 0$ in the new policy, and utilize the extra time for the processing times of the task(s) in between $k$ and $\ell$. Thus, the new policy will give strictly higher reward. We can continue to apply this procedure until either $r_{k}' = 0$ or $x_{\ell-1}' = x_{max}$. If $r_{k}' = 0$, then we determine a new $k$ among the remaining tasks with the highest index $i$ such that $r_i \neq 0$ and $x_i\neq x_{max}$. Then, we apply the same procedure: If $x_{\ell-1}' = x_{max}$, we choose the smallest task index with $\ell>k$ and  $x_\ell = x_{max}$. We continue to apply this procedure until $r_i = 0$ for all $i$ such that $x_i\neq x_{max}$.
\end{Proof}

\begin{figure}[t]
\centerline{\includegraphics[width=0.9\columnwidth]{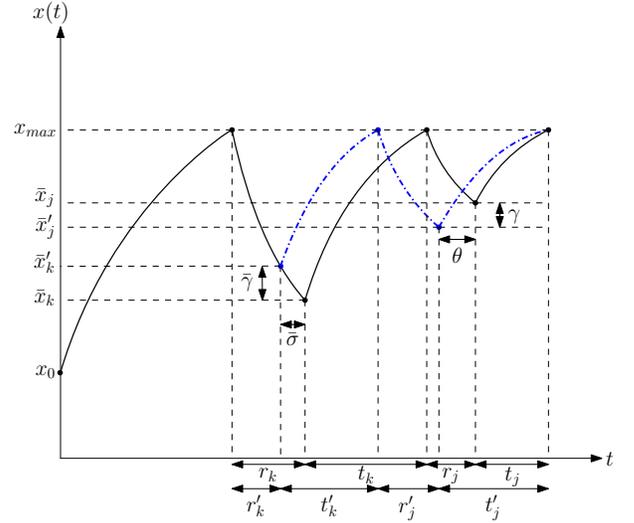}}
		\caption{Black curve shows the hypothetical optimal policy; blue dashed curve shows the new feasible policy that yields a larger reward (Lemma~\ref{lemma_4}).}
		\label{lemma4_fig}
\vspace*{-0.4cm}
\end{figure}

Lemma~\ref{lemma_3} implies that, for a given $i$, if $r_i\neq 0$, then $x_i=x_{max}$, i.e., if the operator rests during processing a task $i$, then he/she must reach $x_{max}$ at the end of processing that task. This also implies that once the operator reaches $x_{max}$ for the first time, since he/she needs to rest to continue, the utilization ratio should hit the upper bound after processing each and every task after this point on. The next lemma states that the operator should allocate equal amount of time for each task he/she processes after reaching the maximal allowable utilization ratio once.

\begin{lemma}\label{lemma_4}
After the point where the utilization ratio reaches $x_{max}$ for the first time (or the point where processing another task would increase the utilization ratio beyond $x_{max}$), the operator spends equal amount of time for processing each remaining task.
\end{lemma}

\begin{Proof}
After reaching $x_{max}$, the operator needs to idle in order to process another task. From Lemma~\ref{lemma_3}, we know that once the operator idles, his/her utilization ratio needs to reach $x_{max}$ again. Consider tasks $j$ and $k$ where $r_k$ and $r_j$ are both non-zero. Assume for contradiction that $r_j\neq r_k$. Without loss of generality, assume $r_j <r_k$, which also implies $t_j<t_k$. Then, we have $r_j+t_j < r_k+t_k$. Consider a new policy where $r_j' = r_j +\sigma$, $t_j' = t_j+\theta$, $r_k' = r_k -\bar{\sigma}$, and $t_k' = t_k -\bar{\theta}$. We can choose them in such a way that $\sigma + \theta = \bar{\sigma} +\bar{\theta} = \Delta T$. Let $\bar{x}_j$, $\bar{x}_k$, $\bar{x}_j'$, and $\bar{x}_k'$ denote the utilization ratios of tasks $j$ and $k$ for the original and new policies such that $\bar{x}_j>\bar{x}_k$ and $\bar{x}_j'\geq \bar{x}_k'$; see Fig.~\ref{lemma4_fig}. Then, for task $j$,
\begin{align}
\bar{x}_je^{-\alpha \sigma } & =   \bar{x}_j' \\
1-e^{-\alpha  t_j }+\bar{x}_j e^{-\alpha  t_j } & = x_{max}= 1-e^{-\alpha  t_j' }+\bar{x}_j' e^{-\alpha  t_j' } \\
\bar{x}_j  &= \frac{1-e^{-\alpha \theta }}{1-e^{-\alpha  \Delta T } }
\end{align}
Similarly, for task $k$,
\begin{align}
\bar{x}_k'e^{-\alpha \bar{\sigma} } & =   \bar{x}_k \\
1-e^{-\alpha  t_k }+\bar{x}_k e^{-\alpha  t_k }=  x_{max} &=1-e^{-\alpha  t_k' }+\bar{x}_k' e^{-\alpha  t_k' } \\
\bar{x}_k'= \bar{x}_k e^{\alpha \bar{\sigma}}  &= \frac{1-e^{-\alpha \bar{\theta} }}{1-e^{-\alpha \Delta T }}
\end{align}

\begin{figure}[t]
	\centerline{\includegraphics[width=1\columnwidth]{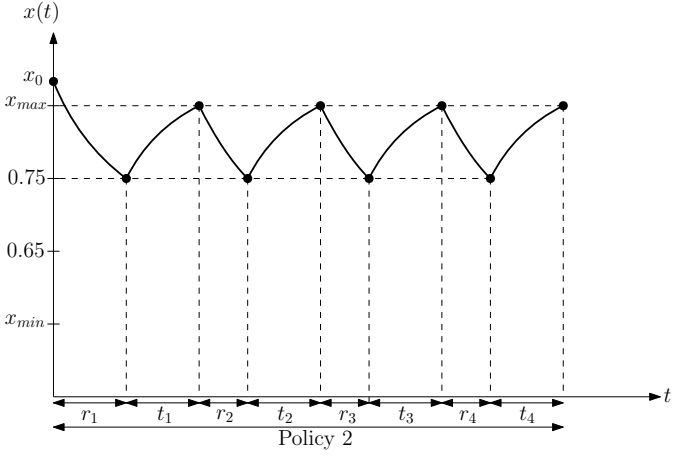}}
	\caption{Evolution of $x(t)$ when $x_0$ is high enough so that the operator needs to rest before processing any tasks.}
	\label{Figure1}
\vspace*{-0.4cm}
\end{figure}

Since $\bar{x}_k <\bar{x}_k'\leq \bar{x}_j'< \bar{x}_j$, we have $ \theta >\bar{\theta}$. Thus, $t_j'+t_k' = t_j+t_k+\theta - \bar{\theta} > t_j+t_k $ due to  $\theta - \bar{\theta} >0$. Also, $t_k'-t_j' = t_k-\bar{\theta}-t_j-\theta < t_k -t_j$. Note that the total task processing time is increased and the difference between time allocations is decreased. This new policy will give strictly larger utility due to the monotonicity and concavity of the utility function $u(t)$. Thus, we reached a contradiction and $r_j\neq r_k$ cannot be optimal.
\end{Proof}

\section{The Optimal Solution}
The optimal solution is composed of two major policies: \emph{Policy 1,} where the operator processes tasks without idling until either he/she reaches $x_{max}$ for the first time or processing another task would force him/her to exceed the allowed $x_{max}$ so he/she needs to stop without reaching $x_{max}$; and \emph{policy 2,} which starts either when the operator reaches $x_{max}$ for the first time, or when processing another task would force him/her to exceed $x_{max}$. After reaching $x_{max}$ for the first time, the operator alternates between resting (idling) and processing tasks in equal amounts. We define $m$ as the number of tasks processed in \emph{policy 1}. We define $\tilde{t}_1$ and $\tilde{t}_2$ to be the processing times for tasks processed in \emph{policy 1} and \emph{policy 2}, respectively. We define $\tilde{r}_1$ to be the idling time right before the operator reaches $x_{max}$ for the first time, and $\tilde{r}_2$ to be the idling time after the operator reaches $x_{max}$. Note that $\tilde{r}_1$ might not always exist. Next, we describe the optimal solution in terms of these.

\begin{figure}[t]
	\centerline{\includegraphics[width=1\columnwidth]{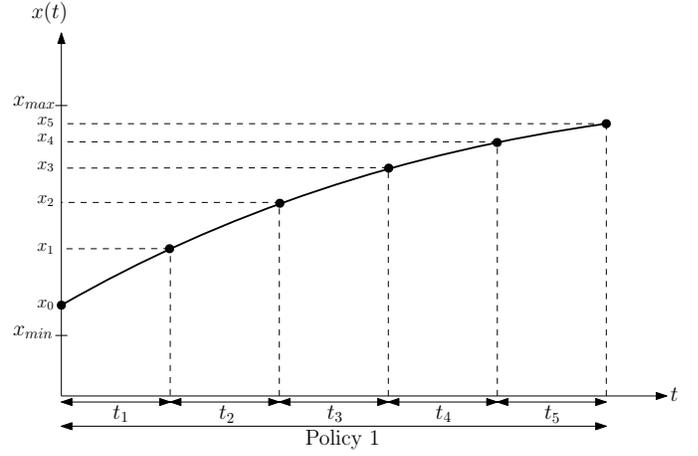}}
	\caption{Evolution of $x(t)$ when $x_0$ is low enough so that the operator does not need to rest for any task.}
	\label{Figure4}
\vspace*{-0.4cm}
\end{figure}

When the operator starts with an initial utilization ratio $x_0$, there are two options: either $x_0$ is high enough that the operator needs to rest before beginning to process any tasks (an example of this is given in Fig.~\ref{Figure1}), or $x_0$ is small enough that, from Lemma~\ref{lemma_3}, the operator processes some number of tasks without idling until he/she reaches $x_{max}$. If $x_0$ is sufficiently small, then the operator can process all of the tasks without any idling as described in Lemma~\ref{lemma_2}. In this special case, $m =N$ which means that all tasks are processed in \emph{policy 1} and $t_i = \tilde{t}_1 =\frac{T}{N}$ and $r_i = 0$ for all tasks. An example of this particular case is shown in Fig.~\ref{Figure4}. In the case when $x_0$ is high, the optimal policy is: $r_1 = \tilde{r}_1$, $r_i =\tilde{r}_2$, $i\in \{2,\dots,N\}$, and $t_i = \tilde{t}_2$, $i\in\{1,\dots,N\}$. Note that for this special case $m = 0$ which means that all the tasks are processed in \emph{policy 2}. An example of this particular case is shown in Fig.~\ref{Figure1}.

\begin{figure}[t]
	\centerline{\includegraphics[width=0.97\columnwidth ]{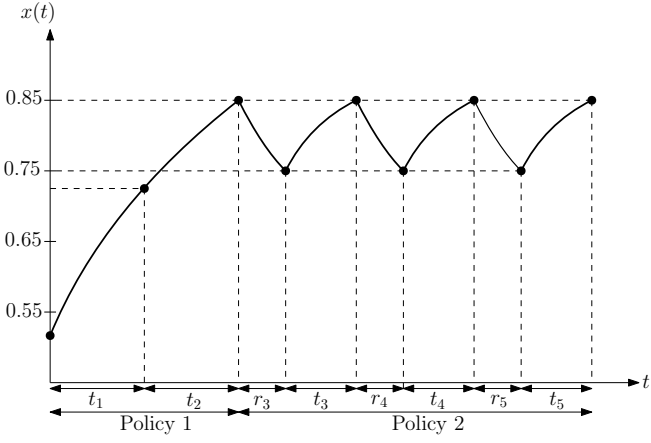}}
	\caption{Evolution of $x(t)$. Here $N=5$ and $m=2$. The operator processes $m=2$ tasks before he/she hits $x_{max}$.}
	\label{figure2}	
\vspace*{-0.4cm}
\end{figure}

\begin{figure}[t]
	\centerline{\includegraphics[width=0.97\columnwidth ]{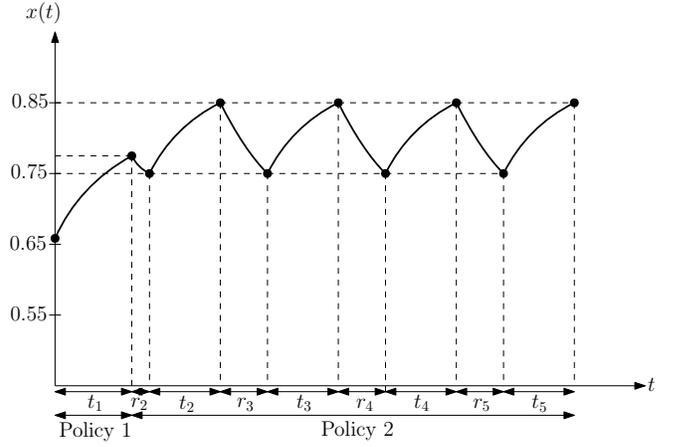}}
	\caption{Evolution of $x(t)$. Here $N=5$ and $m=1$. The operator processes $m=1$ task, and then needs to idle, as processing another task would lead $x(t)$ to exceed $x_{max}$.}
	\label{figure3}
\vspace*{-0.45cm}
\end{figure}

The two cases described above are special cases where all the tasks are processed either in \emph{policy 1} or in \emph{policy 2}. In general, some of the tasks are processed in \emph{policy 1} and the remaining tasks are processed in \emph{policy 2}. These cases correspond to $0<m<N$. For this, there are two possibilities: In the first possibility, the operator can reach $x_{max}$ for the first time without idling. An example of this shown in Fig.~\ref{figure2}. In this case the optimal policy is: $t_i = \tilde{t}_1$, $r_i = 0$, $i\in \{1,\dots,m\}$, and $t_i = \tilde{t}_2$ and $r_i = \tilde{r}_2$, $i\in \{m+1,\dots,N\}$. Note that there is no $\tilde{r}_1$ in this case. In the second possibility, the operator will need to rest just before he/she reaches $x_{max}$ for the first time. An example of this is shown in Fig.~\ref{figure3}. In this case, the optimal solution is: $t_i = \tilde{t}_1$, $r_i = 0$, $i\in \{1,\dots,m\}$ and $t_i = \tilde{t}_2$, $i\in \{m+1,\dots,N\}$ and $r_{m+1} = \tilde{r}_1$ and $r_i =\tilde{r}_2$, $i\in \{m+2,\dots,N\}$. Note that we can determine $\tilde{r}_2$ from $\tilde{t}_2$. Thus, in general, in order to completely characterize the optimal solution, we need to solve for $m$, $\tilde{t}_1$ and $\tilde{t}_2$. In the following lemma, we further characterize $\tilde{t}_1$ and $\tilde{t}_2$.

\begin{lemma}\label{lemma_5}
In the optimal policy, if the evolution of $x(t)$ is as in Fig.~$\ref{figure2}$, then $\tilde{t}_1>\tilde{t}_2$. If it is as in Fig.~$\ref{figure3}$, we can have the following cases: If $N\geq2m$, then $\tilde{t}_1>\tilde{t}_2$. If $N<2m$, then we need to check the relation between $m\frac{1-x_m(\tilde{t}_1,x_0)}{x_m(\tilde{t}_1,x_0)}$ and $(N-m)\frac{1-\bar{x}_{m+1} (\tilde{t}_2) }{\bar{x}_{m+1} (\tilde{t}_2)}$. If $m\frac{1-x_m(\tilde{t}_1,x_0)}{x_m(\tilde{t}_1,x_0)}< (N-m)\frac{1-\bar{x}_{m+1} (\tilde{t}_2) }{\bar{x}_{m+1} (\tilde{t}_2)}$, then $\tilde{t}_1>\tilde{t}_2$. If $m\frac{1-x_m(\tilde{t}_1,x_0)}{x_m(\tilde{t}_1,x_0)}> (N-m)\frac{1-\bar{x}_{m+1} (\tilde{t}_2) }{\bar{x}_{m+1} (\tilde{t}_2)}$, then $\tilde{t}_1<\tilde{t}_2$. If $m\frac{1-x_m(\tilde{t}_1,x_0)}{x_m(\tilde{t}_1,x_0)}= (N-m)\frac{1-\bar{x}_{m+1} (\tilde{t}_2) }{\bar{x}_{m+1} (\tilde{t}_2)}$, then $\tilde{t}_1=\tilde{t}_2$.
\end{lemma}

\section{Numerical Results} \label{sect:examples}
In this section, we give simple numerical examples for the optimal solution. In the first example, we take $T=7$, $N=3$, $x_{max} = 0.85$, $x_{min}=0.4$ and $x_0=0.6$. The optimal policy for this case is to process all the tasks without idling. This example corresponds to the special case described in Lemma~\ref{lemma_2}, where $x_N =0.8332<x_{max}$. Therefore, the optimal policy is to allocate $t_i = \frac{7}{3} = 2.33$ and $r_i = 0$, for all $i$.

In the second example, we take $T=8.8$, $N=3$, $x_{max} = 0.85$, $x_{min}=0.4$ and $x_0=0.7$. The optimal solution is $t_1= t_2=\tilde{t}_1=2.7726$ and $t_3 =\tilde{t}_2= 2.6740$, $r_1=r_2 = 0$, $r_3=\tilde{r}_2 =0.5808$. The evolution of $x(t)$ in this case is as in Fig.~\ref{figure2}, where there is no $\tilde{r}_1$.

In the third example, we take $T=7.4$, $N=3$, $x_{max} = 0.85$, $x_{min}=0.4$ and $x_0=0.7$. The optimal solution is $t_1= t_2=\tilde{t}_1=2.4013$ and $t_3 =\tilde{t}_2= 2.2610$, $r_1=r_2 = 0$, $r_3=\tilde{r}_1 =0.3364$. The evolution of $x(t)$ in this case is as in Fig.~\ref{figure3}, where there is an $\tilde{r}_1$. In the second and third examples, we observe that $\tilde{t}_1>\tilde{t}_2$.

\bibliographystyle{unsrt}
\bibliography{IEEEabrv,myLibrary_shortened}

\end{document}